\pdfoutput=1
\documentclass[prd,groupedaddress,nofootinbib,twocolumn]{revtex4-1}

\usepackage{amsmath}
\usepackage{amssymb}
\usepackage{graphicx}
\usepackage{bbold}
\usepackage{setspace}
\usepackage{array}
\usepackage[left=1.80cm, right=1.80cm, top=1.80cm, bottom=2.00cm]{geometry}
\usepackage{epsfig}
\usepackage{verbatim}   			
\usepackage[subfigure]{graphfig}
\usepackage{epstopdf}
\usepackage{dcolumn}				
\usepackage[normalem]{ulem}         
\usepackage{caption}
\usepackage{color}
\usepackage{tabularx}
\usepackage{braket}
\usepackage{float}
\usepackage{tikz}
\usepackage[percent]{overpic}
\usepackage{subfigure}
\usepackage{mwe}
\usepackage{soul}
\usepackage{ulem}
\usepackage[utf8]{inputenc}
\usepackage{lineno}
\usepackage[flushmargin]{footmisc}
\usepackage{subfig}

\begin{document}
	
\preprint{ABCD}

\title{The Positive-Parity Baryon Spectrum and the Role of Hybrid Baryons}
\author{Tanjib Khan$^{1}$, David Richards$^{2}$, Frank Winter$^{2}$}
\vspace*{-0.5cm}
\affiliation{
  $^{1}$\mbox{Department of Physics, The College of William and Mary, Williamsburg, Virginia, USA.}\\
  $^{2}$\mbox{Thomas Jefferson National Accelerator Facility, Newport News, Virginia, USA.}
}

\vspace{20pt}

\begin{abstract}
We calculate the low-lying spectra for the positive-parity $\Delta$ and $N$ at two pion masses of $358$ and $278~{\rm MeV}$ using an isotropic clover action with two degenerate light-quark and one strange-quark flavors through the application of the generalized variational method within the distillation framework. The spectrum exhibits the general feature observed in previous calculations using an anistropic clover lattice, with a counting of states at least as rich as the quark model.  Furthermore, we identify states that are hybrid in nature, where gluonic degrees of freedom play a structural role, indicating that such states appear as a feature of the excited baryon spectrum, irrespective of the lattice action, or the precise details of the smearing of the lattice interpolating operators used to identify such states.  
\end{abstract}

\maketitle

\section{Introduction}
	
Lattice QCD (LQCD) provides a powerful numerical approach to solve QCD from the first principles, and has been successfully applied to address a range of key quantities in high-energy and nuclear physics, from the calculation of the ground-state spectrum, to nuclear charges and key measures of hadron structure. The calculation of the excited-state spectrum of QCD presented a particular challenge, in that the formulation of lattice QCD in Euclidean space precludes the direct calculation of scattering amplitudes. 

The starting point for a study of the excited-state spectrum from lattice QCD is the extraction of the discrete, low-lying energy levels on the Euclidean lattice where lattice QCD is formulated.  The most straightforward, albeit na\"{i}ve, approach is to identify those energy levels with the single- or multi-particle states in the spectrum.  Whilst the resulting energy spectrum should be independent of the basis of interpolating operators employed, it has the potential to be highly sensitive to that basis. Thus the first attempts to compute the spectrum employed baryon interpolating operators with a straightforward three-quark structure, mirroring the valence structure \cite{Gockeler:2001db, Sasaki:2001nf, Melnitchouk:2002eg, Brommel:2003jm, Basak:2007kj, Mahbub:2009nr, Bulava:2009jb, Mahbub:2010jz, Bulava:2010yg, Alexandrou:2013fsu}. 

A focus of several studies has been the nature of the Roper resonance, the lowest-lying positive-parity excitation of the $N$, and whether the spectrum computed on the lattice exhibits the ordering of states revealed in nature whereby the Roper is of lower mass than the lowest-lying negative-parity excitation \cite{Guadagnoli:2004wm, Lasscock:2007ce, Mahbub:2010me, Mahbub:2010rm, Sun:2019aem}.  Here there are indications that the inverse ordering observed at the relatively large quark masses where the calculations have been performed, is reversed in the approach to the physical quark masses \cite{Mahbub:2010me, Mahbub:2010rm}.  The need for a faithful representation of chiral symmetry to obtain a low-lying Roper mass has been argued in ref~\cite{Sun:2019aem}, though a calculation employing the overlap and clover fermion actions on the same set of gauge configurations suggests that some other mechanism may be at play~\cite{Virgili:2019shg}.
For a more complete picture, it is necessary to include multi-hadron interpolators that would couple to states such as $N \pi$, $N \sigma$, $\Delta \pi$, $N \rho$, and $N \pi \pi$ that are present in the spectrum.  The calculation of the two-point functions employing such operators is computationally challenging due to the many additional Wick contractions that enter into their construction.  Despite this, considerable progress has been made at including such operators in the analysis \cite{Kiratidis:2016hda, Lang:2016hnn, Andersen:2017una}, though the evidence of the emergence of a low-lying Roper as a $N\pi$ scattering state in these works remains scant. 

The excited states are of course resonances, and a revolution over the last few years has arisen from the realization that the energy shifts at finite volume could be related to infinite-volume scattering amplitudes \cite{Luscher:1985dn} which has transformed our ability to study the resonance spectrum, and the interaction of hadrons, from lattice QCD. The application of these methods to meson spectroscopy has been extensive\footnote{For a recent review, see ref. \cite{Briceno:2017max}}, and the first steps have been taken in applying them to the more computationally demanding baryon spectrum \cite{PhysRevD.88.014511, Andersen:2017una, Lang:2016hnn}. Given the computational demands that such calculations entail, a finite-volume Hamiltonian approach has been developed to relate the energies levels computable in currently practical computations to physical scattering parameters \cite{Wu:2017qve}, and applied to the baryon spectrum in ref. \cite{Liu:2016uzk}.

The focus of our work here is on the higher states in the spectrum. Considerable insight into the excited baryon spectrum of QCD has been obtained, most notably, the extracted spectrum is found to be at least as rich as the quark model \cite{PhysRevD.34.2809, PhysRevD.18.4187, PhysRevD.28.170}, and exhibits a counting of energy levels commensurate with ${\rm SU(6)} \otimes {\rm O(3)}$ spin-flavor symmetry \cite{Edwards:2011jj}. Indeed, the ordering of the states was found to be in agreement with the predictions of the constituent quark model which can be considered as an intermediate phenomenological model consistent with the experimental results.  However, such a model interprets the Roper resonance predominantly as a $qqq$ state augmented by a meson cloud \cite{Burkert:2017djo}, and fails to find the low-mass Roper seen in experiments.

Moreover, the positive-parity excited baryon spectrum reveals suggestions of ``hybrid'' states, that is those in which the gluonic degrees of freedom play an essential, structural role, beyond those of the quark model, with a common mechanism with comparable states in the meson section \cite{Dudek:2012ag}. For the case of mesons, there can exist ``exotic'' states that have quantum numbers $J^{PC}$ not available within a regular $q \bar{q}$ valence structure. Such states may have a dominant $qq\bar{q}\bar{q}$ component, so-called tetraquarks \cite{Cheung:2017tnt, Junnarkar:2018twb}, or be predominantly ``hybrid'' $q\bar{q} g$ states \cite{CHANOWITZ1983211, Dudek:2009qf, Dudek:2010wm, Dudek:2011tt, Dudek:2013yja}, with a manifestation of the gluonic valence component. Thus it is straightforward to separate an ``exotic'' meson state from a regular one.  But for baryons, the regular $qqq$ states can have all the $J^{P}$ values. So, hybrid or so-called pentaquark states will always have to``share" quantum numbers with regular states, thus making them very difficult to identify.

There has been a number of models proposed to calculate the spectrum of hybrid baryons. There is the bag model \cite{BARNES198389} where the quark and gluon fields are confined within a cavity with the fields satisfying appropriate boundary conditions at the wall of the cavity. In the flux-tube model \cite{ISGUR1983247, PhysRevD.31.2910}, the quarks sit at the ends of a string-like structure. A meson contains a single flux tube between the quark and the anti-quark; in a hybrid meson, this string is excited by a transverse oscillation. For the case of a baryon, there are three tubes which either meet at a junction or form a triangle. There are also QCD sum-rule methods \cite{PhysRevD.51.R5986} and quark potential models \cite{PhysRevD.28.160} that make predictions about hybrid baryons.

For the case of hybrid baryons, and indeed for the non-exotic hybrids catalogued from lattice calculations in the meson sector \cite{Dudek:2011bn}, their identification has proceeded through observing that the dominant interpolating operators are``hybrid" in  nature, in the sense that the operators vanish for trivial unit gauge configurations. Such an identification by its nature introduces a degree of model dependence, and therefore it is important to study the robustness of such an identification. The aim of this paper is precisely to test such robustness by performing a calculation of the low-lying positive-parity baryon spectrum at pion masses lower than those of ref. \cite{Dudek:2012ag}, using different gauge and fermion actions at a finer spatial lattice spacing.

The remainder of the paper is organized as follows. In section \ref{sec:comp}, we describe the baryon interpolating operators used in our calculation, and briefly outline the distillation methodology used to construct the correlation functions and our implementation of the variational method. Section \ref{sec:comp_details} contains details of our calculation, beginning with the parameters of the ensembles used in the calculation, a description of our fitting procedure, the robustness of the spin identification on our lattices and the stability of the fits under the variation of the parameters of distillation process.  Section~\ref{sec:results} contains our results for the low-lying positive-parity $\Delta$ and $N$ spectrum, together with the quark-gluon assignment of the states, including those we identify as hybrid baryons.  Section \ref{sec:discussion} contains a comparison between the results presented here, and earlier calculations using the anisotropic lattices \cite{Edwards:2011jj, Dudek:2012ag}. In section \ref{sec:conclusion}, we summarize our work and outline future avenues for research.

\section{Computational Strategy}\label{sec:comp}

Since the focus of our calculation is the low-lying positive-parity spectrum, we employ a basis of interpolating operators that have been found to have the dominant overlaps with those states. The construction of the interpolating operators, and the identification of the operators that couple primarily to the low-lying spectrum, has been described in detail in refs.~\cite{Edwards:2011jj} and \cite{Dudek:2012ag}, so we only summarize the salient elements here. The interpolating operators follow a continuum construction, and are expressed as a product of terms describing the flavor structure, Dirac spin and orbital angular momentum implemented through derivatives:
\[
\Big( \mathrm{B}_{\Sigma_\mathrm{F}} \;\otimes\; \big( \mathrm{S}^{\mathrm{P_\mathrm{S}}} \big)^n_{\Sigma_\mathrm{S}} \;\otimes\; \mathrm{D}^{[d]}_{\mathrm{L}, \;\Sigma_\mathrm{D}} \Big)^{\mathrm{J}}
\]
where $B$, $S$ and $D$ denote the flavor, Dirac spin and orbital angular momentum, $L$, components respectively, and $\Sigma_{F},\Sigma_{S},\Sigma_{D}$ are the corresponding permutation symmetries. The resulting operators are projected through suitable Clebsch-Gordon coefficients to total spin $J$; the label $n$ distinguishes different combinations that have the same spin structure, while the $d$ is the order of the gauge-covariant derivative.

For this work, our basis comprises the non-relativistic operators constructed from the upper components, in a suitable $\gamma$ representation of the Dirac spinors, with up to two covariant derivatives, allowing the operators with up to two units of orbital angular momenta. We also include additional operators containing the commutator of two covariant derivatives acting on the same quark field, corresponding to the chromomagnetic components of the gluonic field-strength tensor and denoted by $\mathrm{D}^{[2]}_{\mathrm{L}=1,\mathrm{M}}$ in the following; it is these operators, which vanish for a unit gauge configuration, that are referred to as ``hybrid'' operators, and for which a dominant overlap with a given state, we treat as the signature of the hybrid nature of that state, as we discuss below. Finally, as the calculations are done on a discretized lattice, the operators are subduced from the continuous Hilbert space onto the different lattice irreps. $H_{g}, G_{1g}~{\rm and}~G_{2g}$, where the subscript $g$ denotes positive parity, such that the correlators constructed using these operators, receive contributions in the forward direction only from the positive-parity states.  As a consequence, for total angular momentum $J = \frac{5}{2}$ and higher, the continuum operators are subduced onto multiple irreps, as detailed in Table~\ref{tab:ops}.
\begin{table}
  \renewcommand{\arraystretch}{1.5}
  \begin{tabular}{crrr}\hline\hline

    J & irrep. (dimension) & No.\ of Ops ($\Delta$) & No.\ of Ops (N)\\[0.7ex]
    \hline
    $\frac{1}{2}$ & $G_1(2)$ & 3 (1) & 7 (2) \\[0.7ex]
    $\frac{3}{2}$ & $H(4)$ & 5 (1) & 7 (2) \\[0.7ex]
	$\frac{5}{2}$ & $G_2(2) \oplus H(4)$ & 2 + 2 & 4 + 4 (1 + 1)\\[0.7ex]
	$\frac{7}{2}$ & $G_1(2) \oplus G_2(2) \oplus H(4)$ & 1 + 1 + 1 & 1 + 1 + 1 \\[0.7ex]
   \hline
    \hline
  \end{tabular}
\caption{The numbers of $\Delta$ and $N$ interpolating operators used in the calculation, together with their subductions onto the irreps.\ of the cubic group. In the final two columns, the number in brackets denotes the number of hybrid operators within each irrep..\label{tab:ops}}
\end{table}

The operators created directly from the fields of the lattice Lagrangian couple to states at all scales, thus making the extraction of the lightest states in the spectrum difficult. In order to solve this problem, a linear operator is applied on the quark fields on appropriate time-slices and operators are built from those ``smeared" fields. In this work, the smearing method used is known as Distillation \cite{Peardon:2009gh}. The distillation operator is defined as:
\begin{equation}
\square_{x y} (t) = \sum_{k = 1}^{N_D} \nu_x^{(k)} (t)\; \nu_y^{(k) \dagger} (t) \hspace{10pt} \Rightarrow \hspace{10pt} \square (t) \equiv V_D(t)\; V_D^\dagger (t) 
\end{equation}
where $V_D(t)$ is a $M \times N_D$ matrix, $M = N_c \times N_x \times N_y \times N_z$, $N_c$ is the number of colors, $N_x, N_y, N_z$ are the extents of the lattice in the three spatial directions. The $k^{th}$ column of $V_D(t)$, $\nu_x^{(k)} (t)$ is the $k^{th}$ eigenvector of the second-order three-dimensional differential operator, $\bigtriangledown^2$ evaluated on the background of the spatial gauge fields of time-slice $t$, once the eigenvectors have been sorted by the ascending order of the eigenvalues; $N_D$ is the dimension of the distillation space.  

The reasons for adopting distillation in our calculation are two-fold.  First, the computationally demanding parallel transporters of the theory, the perambulators, depend only on the gauge field, and not on the interpolating operators. So, we can calculate the perambulators on an ensemble of gauge field once, and then reuse them for an arbitrary basis of operators at both the source and the sink, and indeed for a range of calculations. Second, the method provides a more complete sampling of each gauge configuration through imposing a momentum projection at both the source and the sink correlation time-slices. The expectation is that $N_D$ should scale as the physical volume, and the cost of computing the corresponding correlation functions scales as $N_D^4$ for the case of baryons.  Thus there is a computational imperative to use as small a distillation space as possible while still providing a faithful description of the physics, and we will investigate the sensitivity of our results to $N_D$ below. An approach that aims to overcome this scaling with the spatial volume is to stochastically sample the eigenvector space, a method known as LaPH \cite{Morningstar:2011ka}, which we do not pursue here.

A variety of approaches have been developed to obtain the energies and operator overlaps for the excited baryon spectrum~\cite{Allton:1993wc, Mahbub:2009nr}. Here we use the variational method as implemented in ref. \cite{Dudek:2007wv}. Our starting point is the generalized eigenvalue equation (GEV) for the two-point correlator matrix $C(t)$ with elements
\begin{equation}
C_{ij}(t) = \langle 0 \mid O_i(t) \bar{O}_j(0) \mid 0 \rangle\label{eq:gev}
\end{equation}
where without loss of generality we take the source interpolating operator to be at time slice $t=0$, and where $i,j$ label the operators in a given representation of the cubic group. The GEV equation is expressed as
\begin{equation}
C(t)\; u_\alpha = \lambda_\alpha (t,t_0)\; C(t_0)\; u_\alpha
\end{equation}
where $u_\alpha$ are the generalized eigenvectors which satisfy the orthonormality condition $u^\dagger_\alpha\, C(t_0)\, u_\beta = \delta_{\alpha \beta}$, and the corresponding principle correlators are $\lambda_\alpha(t,t_0)$ behaving as
\begin{equation}
\lambda_\alpha (t, t_0) = e^{- m_\alpha (t - t_0)} \Big[ 1+ {\cal O}
  \big( e^{ -\delta m (t - t_0)} \big) \Big]\label{eq:principle}
\end{equation}
Here $m_\alpha$ is the energy of the state labeled by $\alpha$ and $\delta m $ represents the contributions from other states. Our subsequent results are derived from two-state fits to the principle correlators of this form. Furthermore, $C_{ij}(t)$ can be decomposed into the form,
\begin{equation}
C_{ij} (t) = \sum_\alpha \frac{Z_i^{\alpha^*} \; Z_j^\alpha}{2 m_\alpha} e^{-m_\alpha t}
\end{equation}
where the overlap factor $Z_i^\alpha = \langle 0 | O_i | \alpha \rangle$ can be written as, 	
\begin{equation}
Z_i^\alpha = \big( U^{-1} \big)_i^\alpha \; \sqrt{2 m_\alpha}\;\; \mathrm{exp}\bigg( {\frac{m_\alpha t_0}{2}} \bigg).\label{eq:overlap}
\end{equation}
The matrix $U$ is formed using the generalized eigenvectors $u_\alpha$ as its columns. The overlaps can thereby be obtained from the solution of the generalized eigenvector matrix.

\section{Computational Details}\label{sec:comp_details}
Earlier calculations using this basis of operators were performed on anisotropic clover lattices, with a spatial lattice spacing of around $a \simeq 0.123~{\rm fm}$, and an anisotropy, $\xi \equiv a_s/a_t \simeq 3.5$ \cite{Lin:2008pr, Edwards:2008ja} with two mass-degenerate light-quark flavors and a strange-quark. Here we use an isotropic clover action at a smaller lattice spacing $a \simeq 0.094~{\rm fm}$, determined using the $w_0$ scale \cite{Borsanyi:2012zs}, and likewise with $2 + 1$ flavors. We use ensembles at two values of the light-quark masses, corresponding to pion masses of $m_\pi = 358~{\rm and}~278~{\rm MeV}$ respectively. Details of the parameters of our ensemble are listed in Table~\ref{tab:latt}. All the gauge-links entering in the operator constructions are stout-smeared \cite{Morningstar:2003gk}. In order to achieve the best possible sampling of the lattice, we evaluate the two-point correlators from each time-slice on the lattice, and on each configuration, average the correlators over the different time sources to account for the correlations along the temporal direction of the lattices.  We compute the perambulators for $N_D = 64$ eigenvectors.
	
\begin{table}
  \renewcommand{\arraystretch}{1.5}
  \begin{tabular}{ccccc}\hline\hline
    ID & $a$ (fm) & $M_{\pi}$ (MeV) & $L^3 \times N_t$ & $N_{\rm cfg}$\\
    \hline
    $a094m358$ & 0.094(1) & 358(3) & $32^3 \times 64$ & 349\\
    $a094m278$ & 0.094(1) & 278(4) & $32^3 \times 64$ & 259\\
    \hline \hline
  \end{tabular}
\caption{The parameters of the ensembles  where the scale is obtained using $w_0$ \cite{Borsanyi:2012zs}, and $N_{\rm cfg}$ is the number of configurations.\label{tab:latt}}
\end{table}

\begin{figure*}[t]
\center{\includegraphics[scale=0.33]{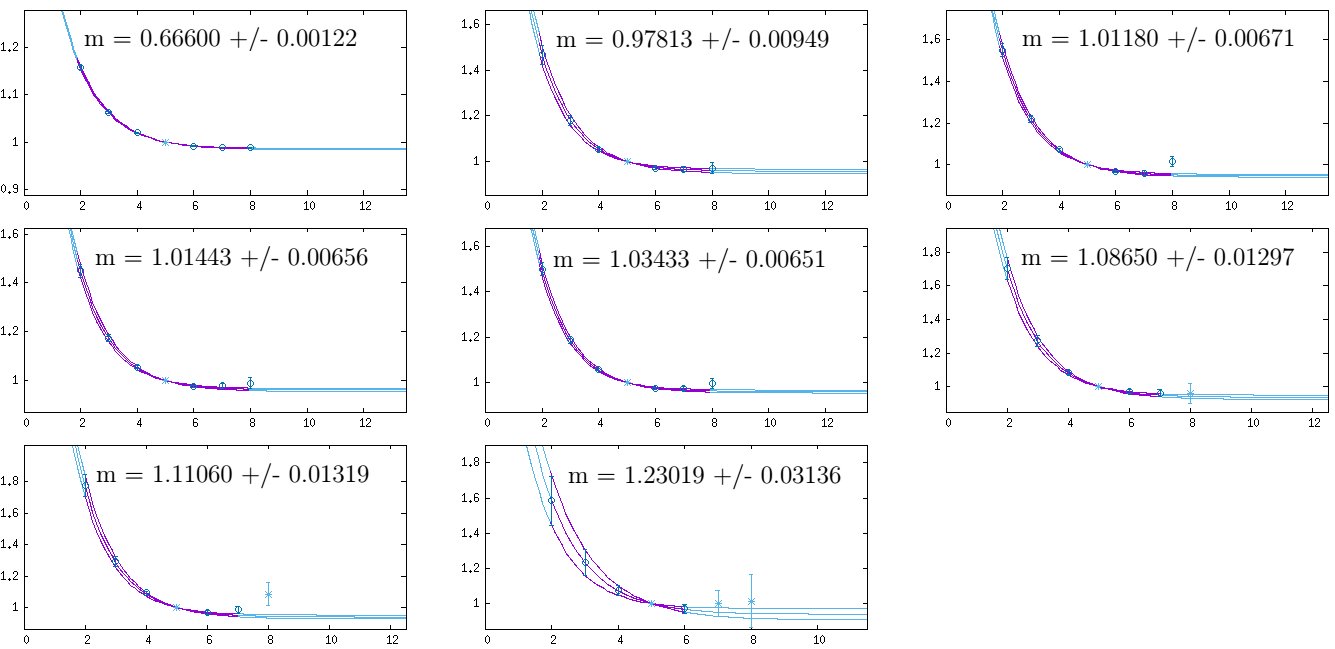}}
\caption{Fits to the principal correlators for the low-lying positive-parity spectrum of the $H_g$ irrep.\ of the $\Delta$ on the ensemble $a094m358$, using $t_0 = 5$. The plots show $\lambda_\alpha (t, t_0).e^{m_\alpha (t - t_0)}$ data on the y-axes and the lattice time-slices on the x-axes; the curves are two-exponential fits as described in the text. In each panel, the mass corresponding to the leading exponential state is labelled by $m$ and given in lattice units. \label{fig:prin_corr}}
\end{figure*}

\subsection{Fitting Procedure}
The fitting procedure we employ to extract the mass spectra and the overlap factors is discussed in detail in ref.~\cite{Dudek:2009qf}, and we only summarize the procedure here. The GEV of eqn.~\ref{eq:gev} is solved over a range of $t_0$, and for each $t_0$, the resulting principle correlators are fit to the two-exponential form,
\begin{equation}
  \lambda_\alpha(t,t_0) = (1-A) e^{-m_\alpha(t-t_0)} + A
    e^{-m'_\alpha(t-t_0)}.\label{eq:fitform}
\end{equation}
We restrict the fitting range such that, for each principle correlator, we only include time-slices for which the noise-to-signal ratio is less than 0.05; in practice, this restricts the largest value of $t$ included in the fits to be around 8. Furthermore, we only include the correlators with source-sink separation greater than two lattice units to avoid possible contact terms. For each $t_0$, our fit to each principle correlator is based on an acceptable $\chi^2/{\rm dof}$.  We also require that the coefficient $A$ in eqn.~\ref{eq:fitform} is less than around 0.1, and that, for a fit to an $N$-dimensional matrix of correlators, the matrix is largely saturated by the lowest-lying $N$ states reflected in that, for each $\alpha$, $m'_\alpha$ is larger than the lowest-lying masses, $m_\alpha$ obtained for all the principle correlators. Finally, we compute the overlap factors of eqn.~\ref{eq:overlap} using the eigenvectors at a reference time-slice, $t_z$. In Figure~\ref{fig:prin_corr}, we show fits to the leading the principal correlators for the $H_g$ irrep.\ of the $\Delta$. For each panel, the curve is the reconstruction from the fitted parameters with the purple region indicating the data points included in the fit. The approach of the plateaux close to unity at large times is indicative of the small value of $A$ in the fits, and the small contribution of the second excited state to each principal correlator.  As anticipated, the uncertainties on the leading mass increase as the mass increases.

\begin{figure*}[t]
\center{\includegraphics[scale=0.20]{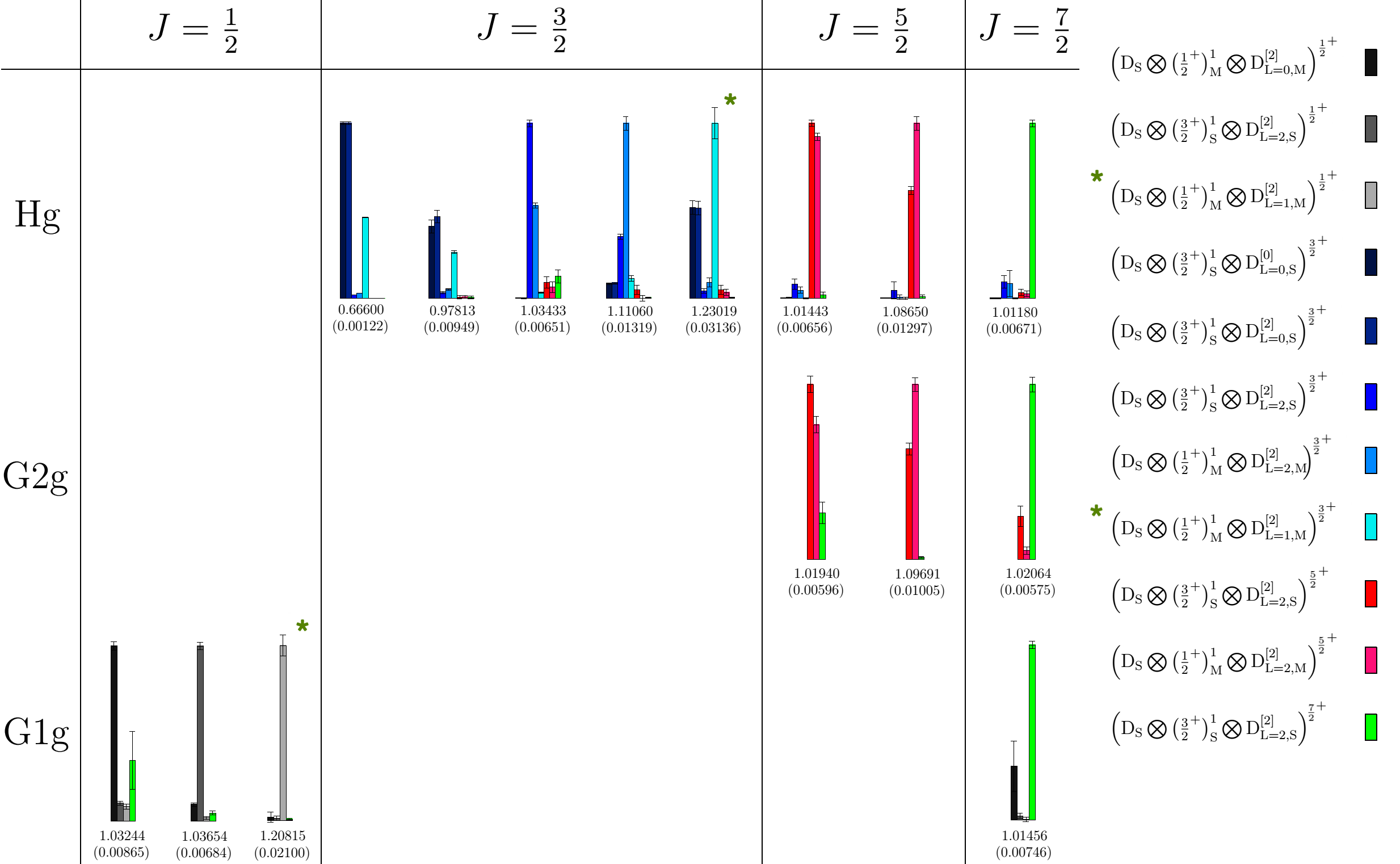}}\caption{Histogram plot of the operator overlaps $Z$ for the $\Delta$ on the $a094m358$ ensemble, normalized such that, for a given operator, the largest overlap across all the states is unity. The overlaps are obtained from a variational analysis across all operators within a given lattice irrep, irrespective of the continuum spins from which they are derived. The histogram plot of each state is accompanied by its mass in lattice units. The asterisks denote the hybrid-type operators, and the energy levels identified with them.
\label{fig:Spin_ID}}
\end{figure*}

\subsection{Spin Identification}
The breaking of rotational symmetry induced by the discretization onto the lattice renders the determination of the spin corresponding to the different energy levels within the irreps. less than straightforward. In the case of the glueball spectrum in pure Yang-Mills theory \cite{PhysRevD.60.034509}, the identification of the spins was accomplished by the identification of degeneracies across different lattice irreps in the approach to the continuum limit.  This requires the generation of ensembles at several lattice spacings, a formidable task once quark degrees of freedom are included, and further requires statistical precision far beyond that attainable with reasonable computational cost to delineate overlapping energies within the spectrum. We need a spin identification method which uses data obtained from only a single lattice spacing, albeit one sufficiently fine that it preserves the rotational symmetry to a sufficient degree at the hadronic scale.

Here we use the method introduced in ref.~\cite{Dudek:2010wm}, and applied for the baryon spectrum in refs.~\cite{Edwards:2011jj,Dudek:2012ag} whereby the operator overlap factors are used to identify the spin of a state. It relies on the observation that each operator used in the calculation carries an essence of the continuum spin of the operator from which it is subduced, and therefore we would expect an operator subduced from, say angular momentum $J$, to have large overlaps only with states of the same continuum angular momentum $J$. Positive-parity states corresponding to the continuum angular momentum $J=\frac{5}{2}$ and $J=\frac{7}{2}$ will appear in the spectrum of the $H_g$ and $G_{2g}$ irreps, and of the $H_g$, $G_{1g}$ and $G_{2g}$ irreps respectively, and we would expect overlaps to be dominated by the operators subduced from the same continuum operators across those irreps. This is indeed what we observe, as can be seen in Figure~\ref{fig:Spin_ID} for the $\Delta$ spectrum, where the overlaps are obtained from a variational analysis using all the operators within a given lattice irrep. Further, we find the resulting energies are degenerate, with, for states of spin $\frac{5}{2}$ and $\frac{7}{2}$, the energies obtained in the $H_g$ irrep. within 1\% of the values obtained in the $G_{1g}$ and, for the case of spin $\frac{7}{2}$, $G_{2g}$ irreps.

\begin{figure*}[t]
\center{\includegraphics[scale=0.40]{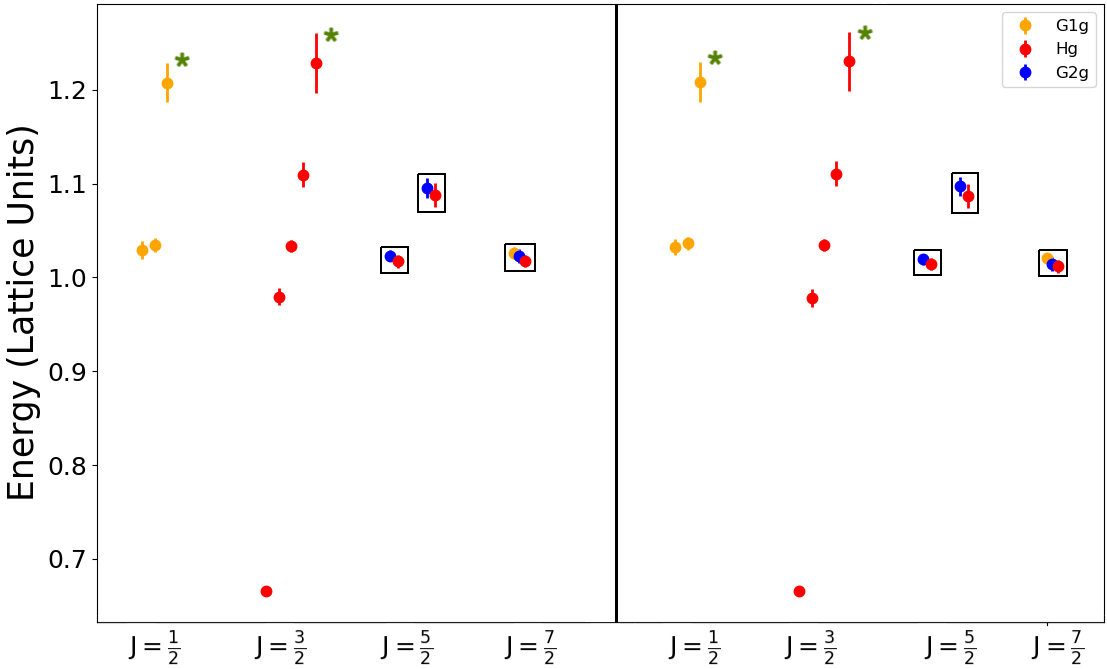}
\caption{Comparison of low-lying $\Delta$ spectra on the $a094m358$ ensemble between fitting only those operators within an irrep.\ derived from the continuum operators of a given angular momentum (left), and fitting all the operators within a lattice irrep., irrespective of their continuum antecedents (right). For the states identified as spin $\frac{5}{2}$ and $\frac{7}{2}$, the boxes contain the energy levels obtained after the subduction onto the different lattice irreps. Energy levels identified as those of hybrid states are denoted by the green asterisks.\label{fig:ang_mom}}}
\end{figure*}
Rather than applying the variational method to a basis comprising all the operators within a lattice irrep., we instead apply the method to a more restricted basis of operators comprising those operators within an irrep.\ derived from a given continuum $J$. In Figure~\ref{fig:ang_mom}, the $\Delta$ spectrum obtained by analysing all the operators within a given lattice irrep. is compared with that where we apply the varational method in each lattice irrep. to only those operators derived from a given continuum spin. The comparison reveals that there are no significant differences between these two spectra, prompting us to analyse the operators of each angular momentum separately as this requires calculating the two-point correlators using a smaller basis of operators at one time; since the computational cost of computing the full correlation matrix goes as the square of the operator basis, this reduces the computational cost significantly.

\subsection{Stability under variation of Distillation Space}
The previous studies of the low-lying baryon spectrum using this implementation of distillation employed $N_D = 56$ distillation eigenvectors, on a $16^3$ spatial lattice with $a \simeq 0.123~{\rm fm}$. With the expected scaling of the number of eigenvectors with physical volume discussed earlier, that would suggest that as many as 230 eigenvectors might be needed to capture the same physics on the ensembles employed here, with in excess of three times the physical spatial volume.  In this paper, we have generated perambulators and the baryon elementals that encode the operators for $N_D=64$ eigenvectors, and begin our discussion by examining the sensitivity of the extracted spectra to the variation of $N_D$.  A study of the various charges of the $N$, both for a state at rest \cite{Egerer:2018xgu}, and at non-zero spatial momentum \cite{Egerer:2020hnc}, performed on the same lattices as those employed here, suggests that the ground-state properties can indeed be resolved with this number of eigenvectors.  However, ascertaining the sensitivity of our results for the spectrum and overlaps of the excited states is an important prerequisite for our subsequent discussion.  

\begin{figure*}[t]
\center{\includegraphics[scale=0.33]{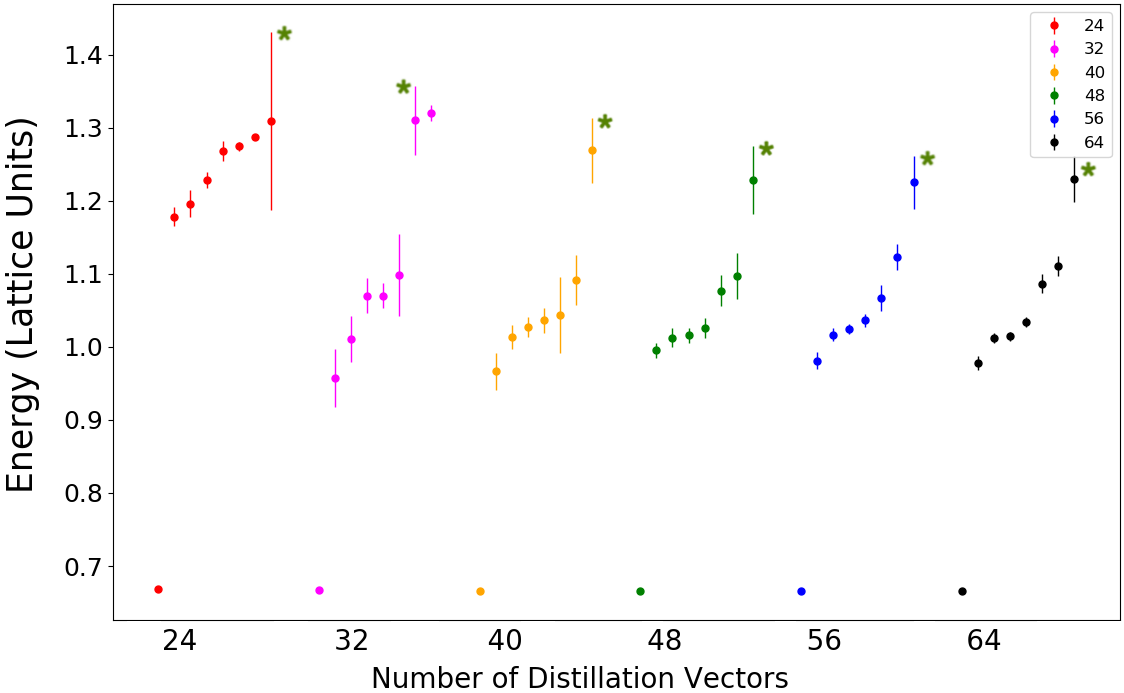}}\caption{The dependence of the $\Delta$ spectrum in the $H_g$ irrep.\ on the number of distillation vectors $N_D$ for the ensemble $a094m358$ of Table~\ref{tab:latt}. The states we identify as hybrid baryons are indicated by the green asterisks.\label{fig:dist}}
\end{figure*}
\begin{figure*}[t]
\centering
\includegraphics[scale=0.315]{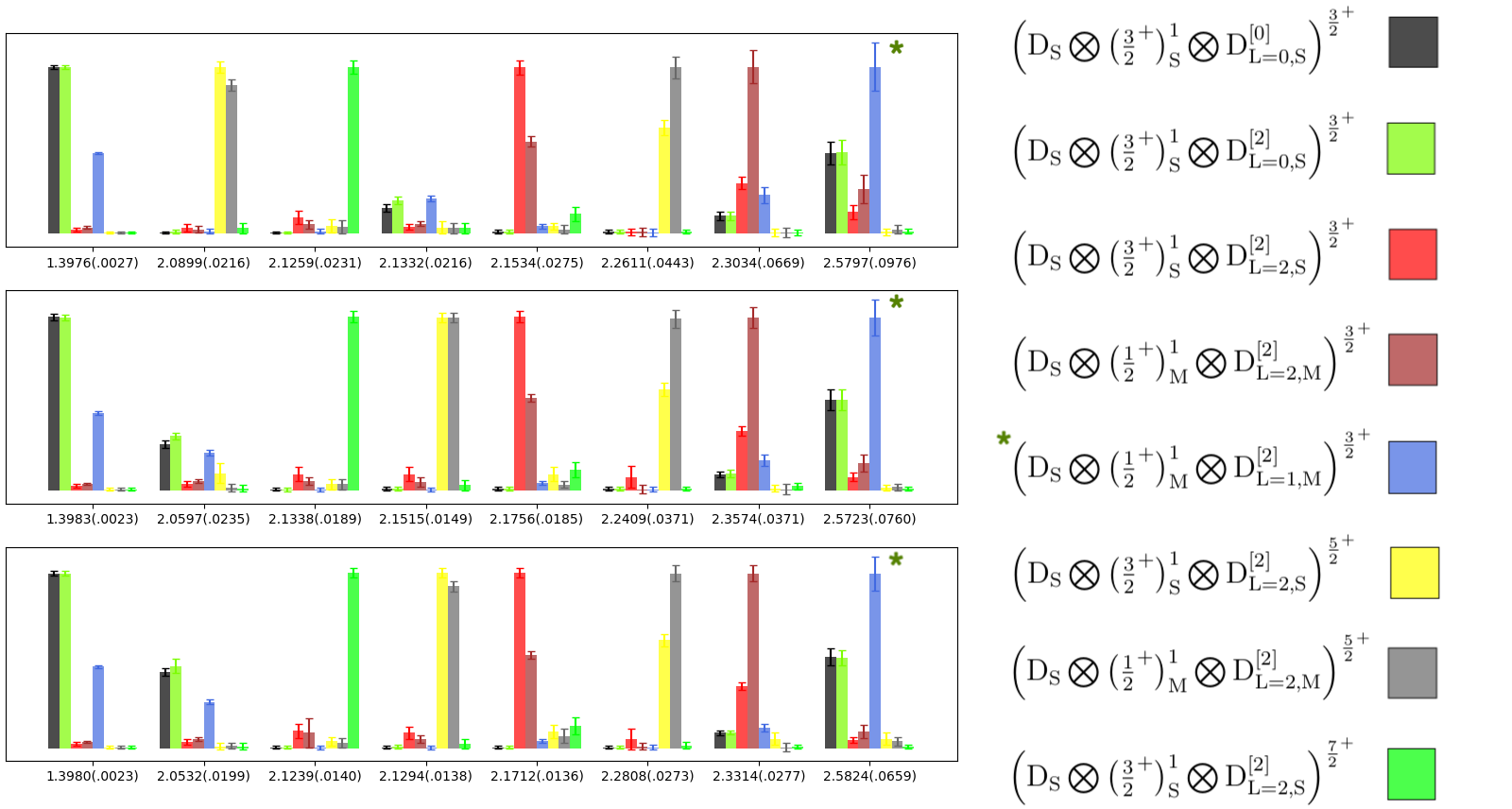}
\caption{The top, middle, and bottom panels show the overlaps of the different operators within the $H_g$ irreducible representation of the $\Delta$ using the $a094m358$ ensemble for $N_D = 48, 56~{\rm and}~64$, respectively. The masses of the states are given in units of GeV.  The asterisks denote hybrid-type operators, and the energy levels identified with them.}
\label{fig:delta_overlap_Nd}
\end{figure*}

In Figure~\ref{fig:dist}, we show the lowest energy levels in the positive-parity $H_g$ irreducible representation of the $\Delta$ as we reduce the number of eigenvectors down to $N_D = 24$.  While the ground state is indeed reliably extracted with only a minimal number of eigenvectors, it is only when we reach $N_D = 56$ eigenvectors that the lowest five states are obtained with acceptable uncertainties, with consistency between the $N_D = 56$ and $N_D = 64$ determinations.  Since a major aim of this work is establishing evidence for the properties of the extracted states, as evidence through the operator overlaps, we likewise need to ensure that these features are robust under the rank of the distillation space. In Figure~\ref{fig:delta_overlap_Nd}, we show the operator overlaps corresponding to the energies in Figure~\ref{fig:dist}; as for the case of the energies, the overlaps, and in particular the dominant operators corresponding to each state, show stability between $N_D = 56$ and $N_D = 64$, but with some qualitative differences, notably in the ordering of the states, for $N_D = 48$. We will therefore use $N_D=64$ in the remainder of this paper.

\section{Results}\label{sec:results}
\begin{figure*}[t]
\center{\includegraphics[scale=0.40]{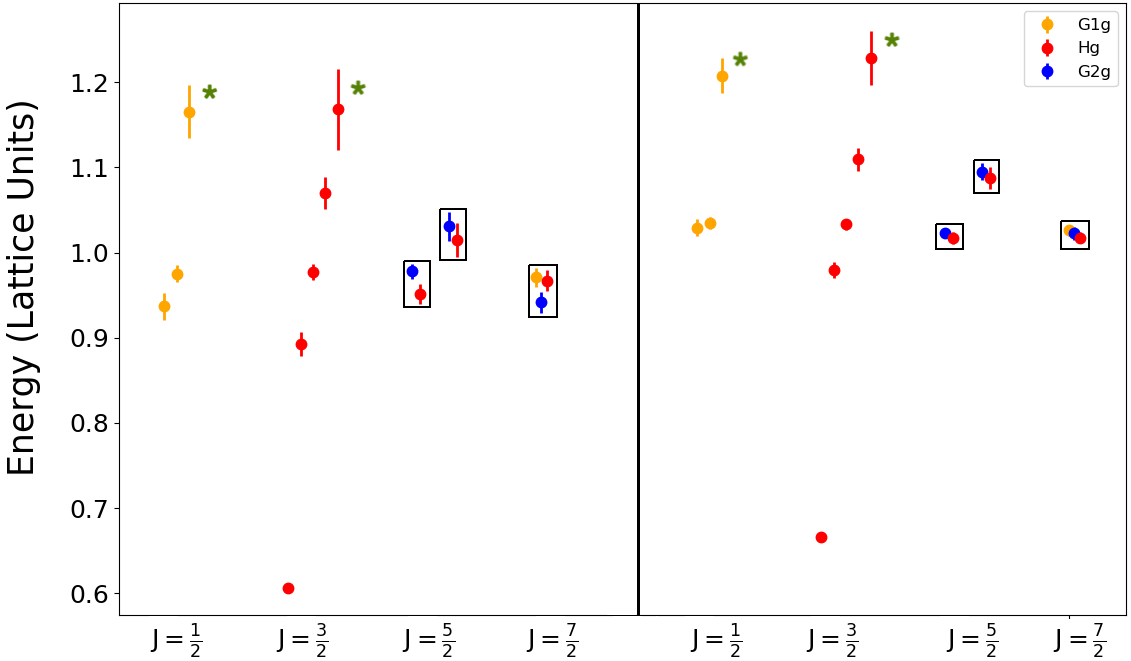}
\caption{The low-lying positive-parity $\Delta$ spectrum in lattice units on the $a094m278$ (left) and $a094m358$ (right) ensembles, using the fitting procedure described in the text. For the states identified as spin $\frac{5}{2}$ and $\frac{7}{2}$, the boxes contain the energy levels obtained after the subduction onto the different lattice irreps. Energy levels identified as those of hybrid baryons are denoted by the green asterisks.\label{fig:delta}}}
\end{figure*}

\begin{figure*}[t]
\center{\includegraphics[scale=0.40]{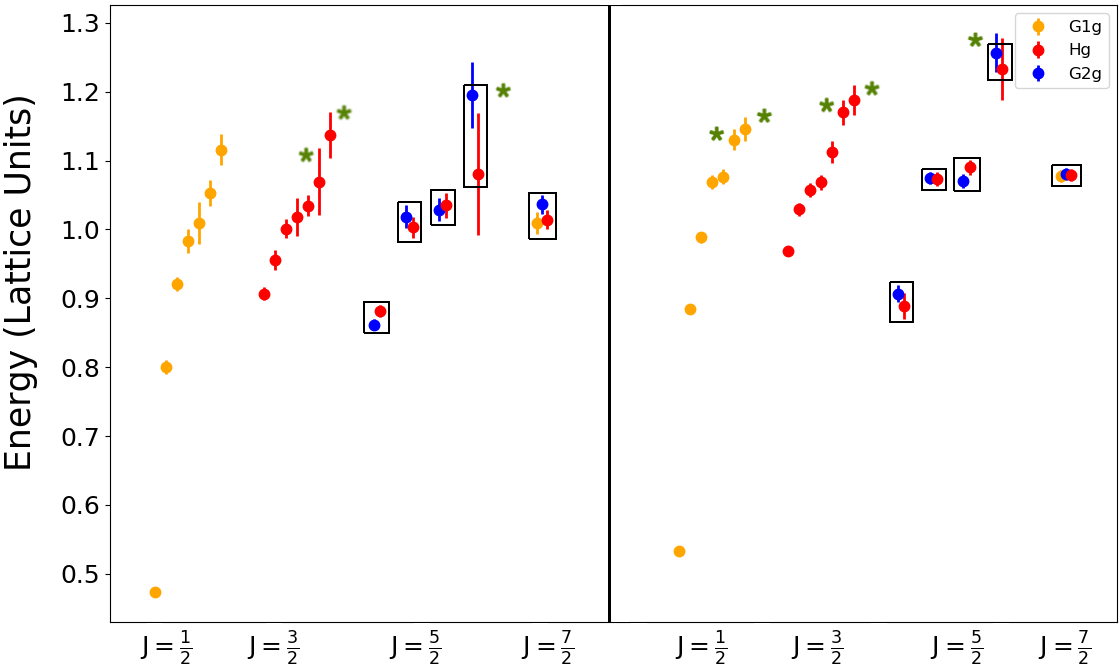}
\caption{The low-lying positive-parity $N$ spectrum in lattice units on the $a094m278$ (left) and $a094m358$ (right) ensembles, using the fitting procedure described in the text. For the states identified as spin $\frac{5}{2}$ and $\frac{7}{2}$, the boxes contain the energy levels obtained after the subduction onto the different lattice irreps. Energy levels identified as those of hybrid baryons are denoted by the green asterisks.\label{fig:nucleon}}}
\end{figure*}

The low-lying positive-parity spectra of the $\Delta$ and $N$ for both the $a094m278$ and $a094m358$ ensembles using this fitting procedure are shown in Figures~\ref{fig:delta} and \ref{fig:nucleon}, respectively. For the spin $\frac{5}{2}$ and spin $\frac{7}{2}$ energy levels, the splittings between the values obtained in the $H_g$ and $G_{2g}$, and in the $H_g$, $G_{2g}$ and $G_{1g}$ irreps, respectively, are remarkably small, reflecting the partial ${\cal O}(a^2)$ breaking of rotational symmetry, and the smaller spatial lattice spacing than that used in comparable studies using an anisotropic lattice. As expected, the quality of the spectrum is somewhat worse at the lighter value of the quark mass, and the spin identification procedure less convincing for the highest states in the $N$ spectrum. We emphasise that the qualitative properties of the spectrum, and in particular the counting of states, is consistent with that obtained on the anisotropic lattices at a coarser value of the spatial lattice spacing, but a considerably finer temporal lattice spacing.  Notably, our calculation does not exhibit the low-lying Roper resonance, in accord with calculations using the anisotropic action, and indeed most calculations using a Wilson-type action.

\subsubsection{Hybrid States}
As we noted in the introduction, in contrast to the case of the meson spectrum, ``exotic" baryons cannot be distinguished through their quantum numbers. Therefore, the identification of baryons as "hybrid" in nature inevitably involves a degree of interpretation. Here we identify the hybrid states as those whose overlap, defined through eqn.~\ref{eq:overlap}, is dominated by the hybrid-type operators, that is those that would vanish for the case of a trivial gauge configuration \cite{Dudek:2012ag}. For the case of the $\Delta$, this identification is very apparent, as can be seen in Figure~\ref{fig:Spin_ID} for the $a094m358$ ensemble, where we find one hybrid state in the $J=\frac{1}{2}$ channel and one in $J=\frac{3}{2}$ channel. For the $N$ spectrum on the $a094m358$ ensemble, we likewise find clear evidence for hybrid-baryon states through the nature of their overlaps, where we identify two states in the $J=\frac{1}{2}$ channel, two states in the $J=\frac{3}{2}$ channel and one state in the $J=\frac{5}{2}$ channel. On the $a094m278$ ensemble, the identification and multiplicities of the hybrid baryons 
follow those of the $a094m358$ ensemble except for the $J=\frac{1}{2}$ channel in the $N$ spectrum, where there is no obvious candidate for a hybrid baryon using the criterion of the operator overlaps. In spite of this, the multiplicity in both the $\Delta$ and $N$ spectrum confirm the findings in the earlier studies using the anisotropic lattice \cite{Dudek:2012ag}, with a multiplicity of states at least as rich as the quark model, and the presence of additional states that appear to be hybrid in nature.

\section{Discussion}\label{sec:discussion}
\begin{figure*}[t]
\center{\includegraphics[scale=0.25]{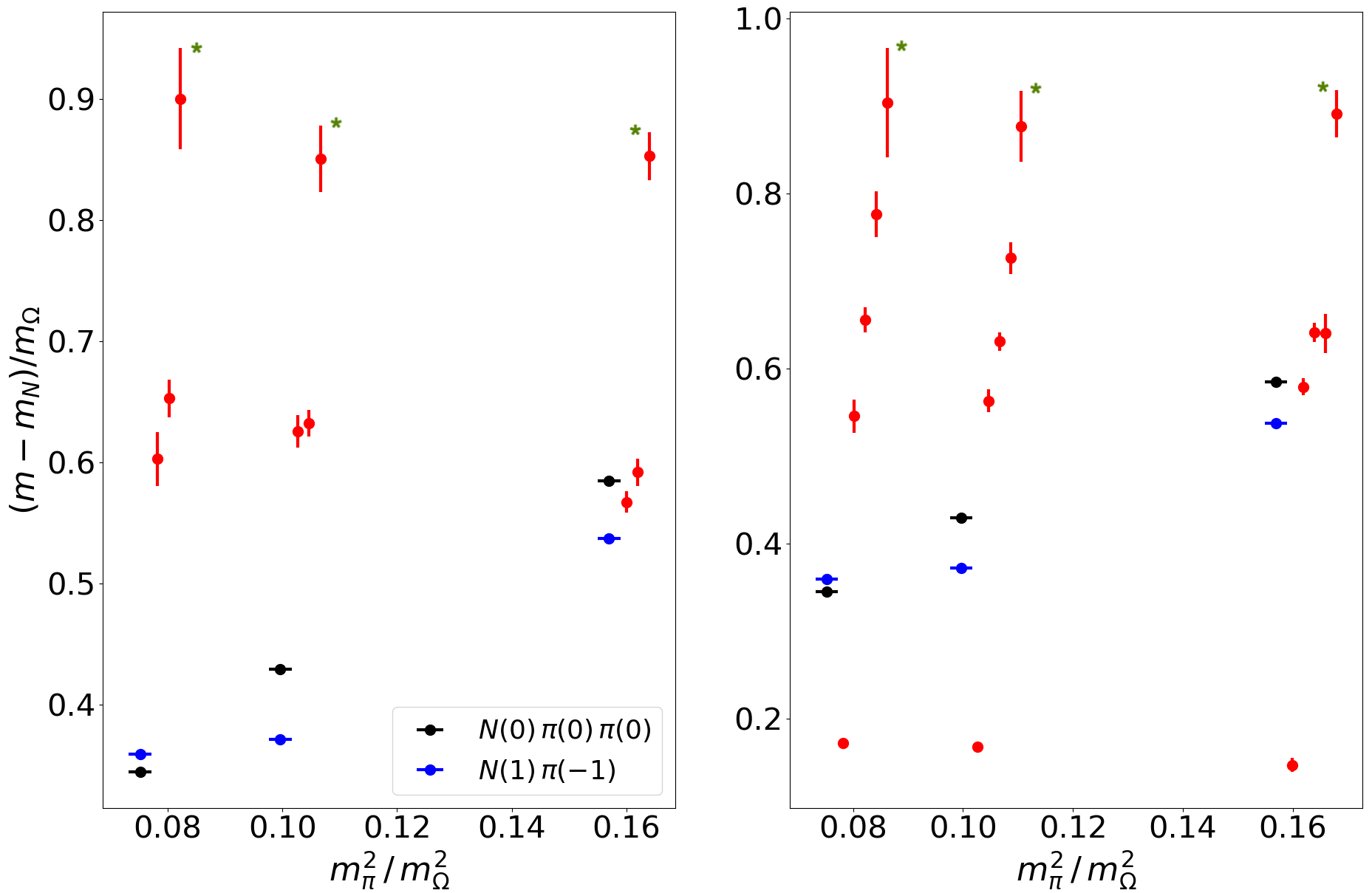}
\caption{The left and right panels show the excitation energies for the $\Delta$ with respect to the ground-state Nucleon on each of our lattices vs.\ $m_\pi^2$, together with the corresponding result from ref.~\cite{Dudek:2012ag}, for the $J=\frac{1}{2}$ and $J=\frac{3}{2}$ channels respectively. We use the $\Omega$ mass to set the scale. The higher excited states are displaced for clarity. The $ N(0) \, \pi(0) \, \pi(0)$ and $N(1) \, \pi(-1)$ energy thresholds are identified by horizontal dashes. \label{fig:delta_comp}}}
\end{figure*}

\begin{figure*}[t]
\center{\includegraphics[scale=0.25]{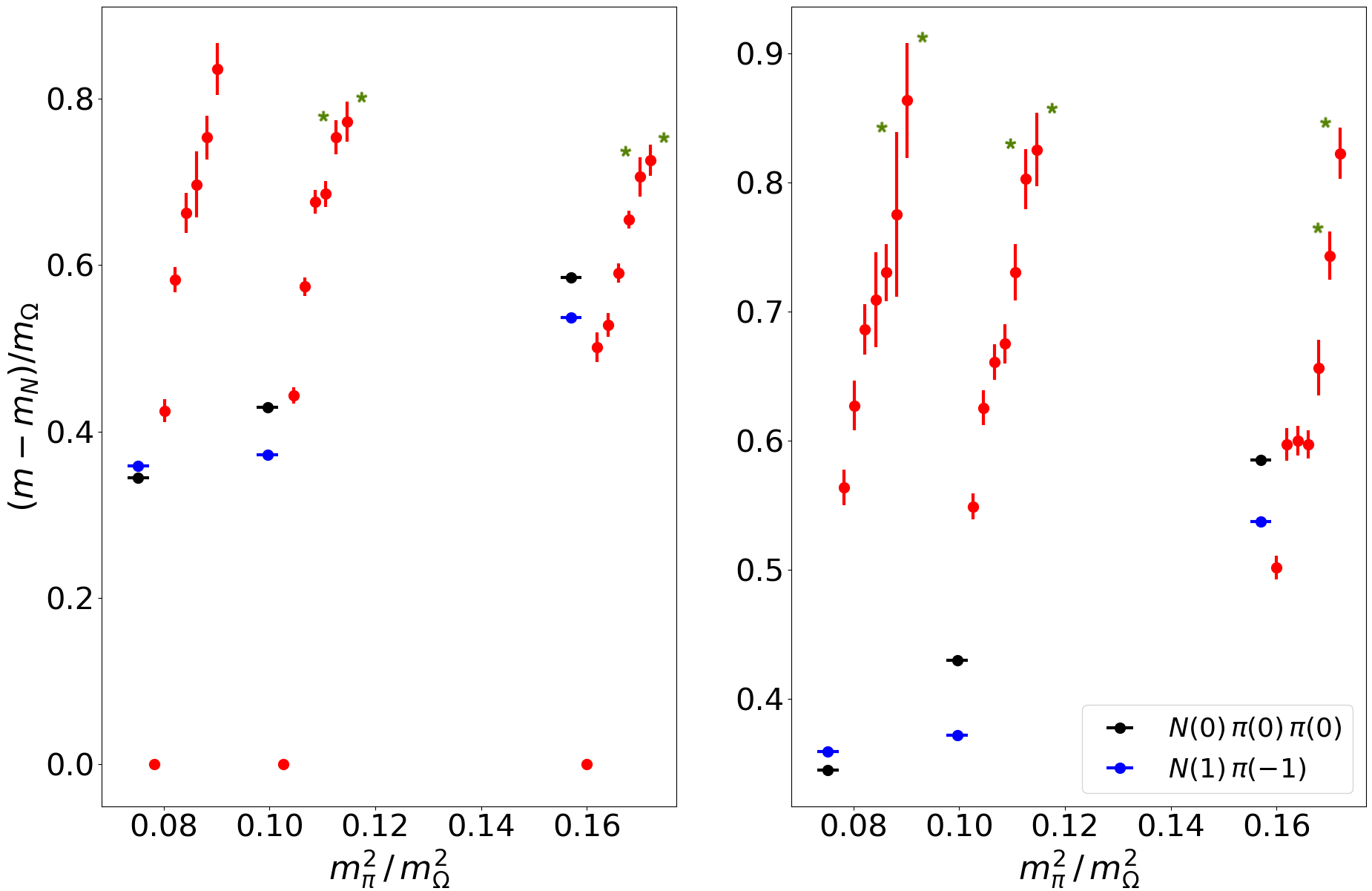}
\caption{The left and right panels show the excitation energies for the $N$ with respect to the ground-state Nucleon on each of our lattices vs.\ $m_\pi^2$, together with the corresponding result from ref.~\cite{Dudek:2012ag}, for the $J=\frac{1}{2}$ and $J=\frac{3}{2}$ channels respectively. We use the $\Omega$ mass to set the scale. The higher excited states are displaced for clarity. The $N(0) \, \pi(0) \, \pi(0)$ and $N(1) \, \pi(-1)$ energy thresholds are identified by horizontal dashes. \label{fig:nucleon_comp}}}
\end{figure*}

We now compare our results with those of previous works, and in particular the previous calculation of the low-lying positive-parity baryon spectrum obtained on the heavier of the two anisotropic clover lattices employed in ref.~\cite{Dudek:2012ag}. To facilitate this comparison, we consider the excitation energy with respect to the ground-state Nucleon mass, in units of the $\Omega$ mass, a quantity that is somewhat insensitive to the light-quark masses. In Figures \ref{fig:delta_comp} and \ref{fig:nucleon_comp}, we show the comparison among these lattices for the $J=\frac{1}{2}$ and $J=\frac{3}{2}$ channels for the $\Delta$ and $N$, respectively.  Also shown are the lowest-lying non-interacting two-particle energy levels.

A notable feature of most states for both the $\Delta$ and $N$ is that the splitting with respect to the ground-state Nucleon mass shows only a weak dependence on the quark masses, while the energies of the non-interacting two-particle states exhibit a far stronger dependence. This suggests that we are observing predominantly ``single-hadron" states rather than multi-hadron states, and leads further support to our assertion that the three-quark operators used in this study couple only weakly to the multi-hadron states. These observations are more prominent for the hybrid baryons whose masses, with respect to the ground-state Nucleon mass, remain more or less the same irrespective of the quark masses. However, there is one qualification to this observation, namely that the first excited-state energy seen in the $N^{\frac{1}{2}}$ channel exhibits a stronger dependence on the light-quark masses, and is indeed consistent with that of $\; N(0) \, \pi(0) \, \pi(0)$ and $N(1) \, \pi(-1)$ multi-hadron states.

\begin{figure*}[t]
\center{\includegraphics[scale=0.33]{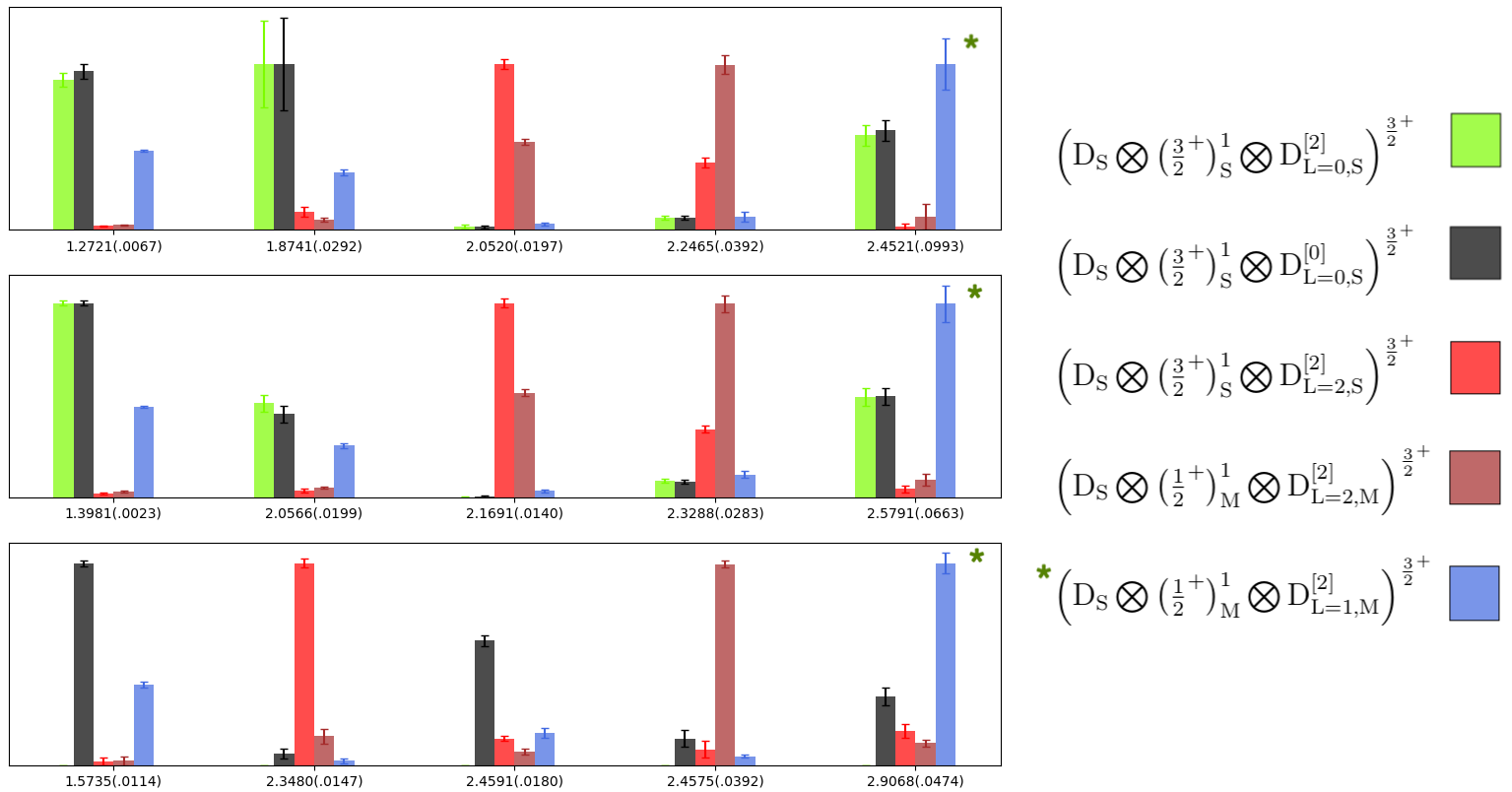}
\caption{The top, middle, and bottom panels show the overlaps of the different operators within the channel J = $\frac{3}{2}$ of the $\Delta$ for the ensembles $a094m278$, $a094m358$ and from \cite{Dudek:2012ag} respectively. The masses of the states are given in the units of GeV. The asterisks denote hybrid-type operators, and the energy levels identified with them.\label{fig:delta_overlap}}}
\end{figure*}

\begin{figure*}[t]
\center{\includegraphics[scale=0.32]{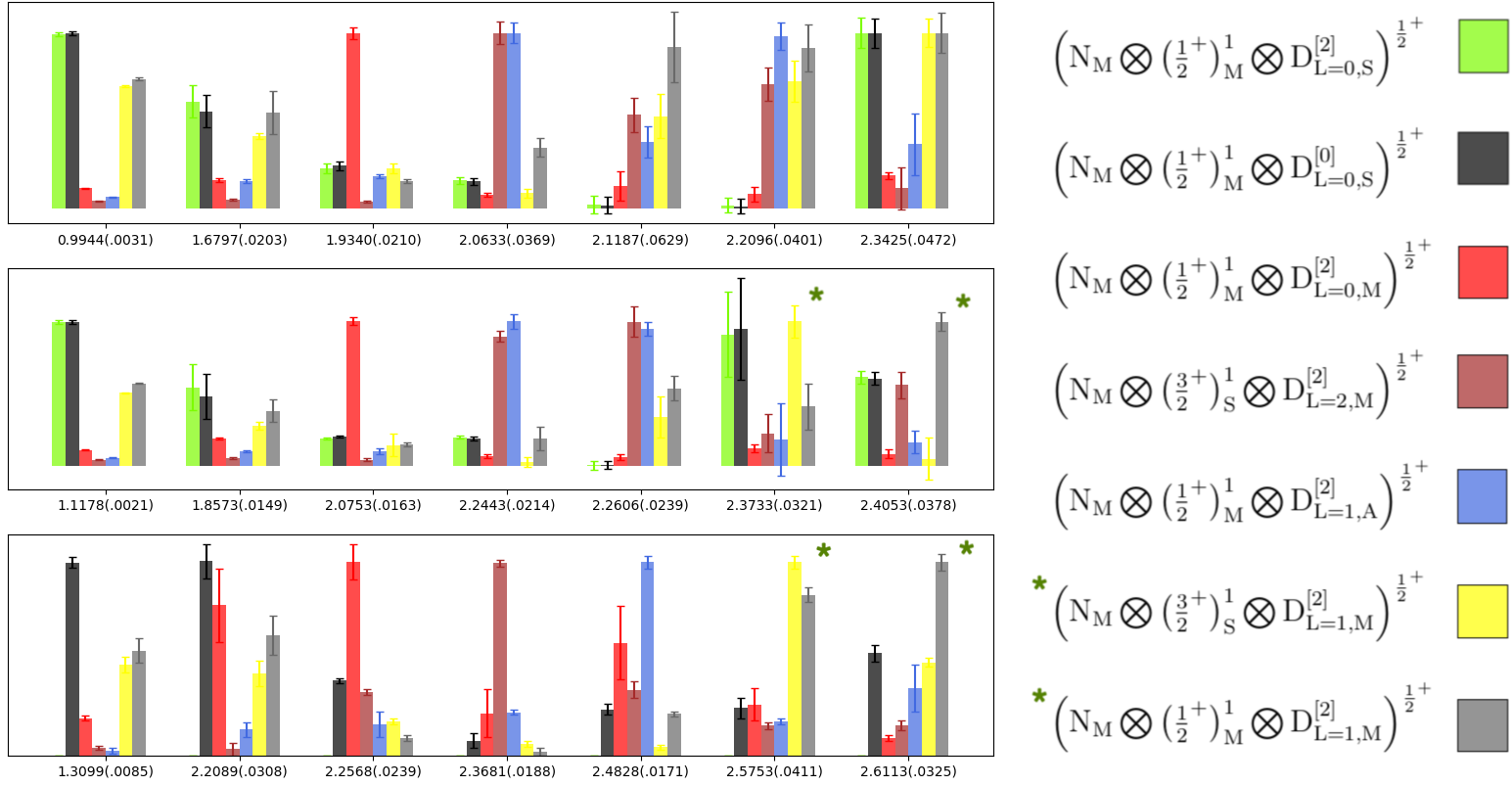}
\caption{The top, middle, and bottom panels show the overlaps of the different operators within the channel J = $\frac{1}{2}$ of the $N$ for the ensembles $a094m278$, $a094m358$ and from \cite{Dudek:2012ag} respectively. The masses of the states are given in the units of GeV. The asterisks denote hybrid-type operators, and the energy levels identified with them.\label{fig:nucleon_overlap}}}
\end{figure*}

A focus of this paper is whether the identification of hybrid baryons is indeed robust. In Figures \ref{fig:delta_overlap} and \ref{fig:nucleon_overlap}, we show that the dominant operators for each of the states of all three ensembles for $J = \frac{3}{2}$ in the case of the $\Delta$, and for $J = \frac{1}{2}$ in the case of the $N$. For the $\Delta$, the hybrid baryon in each of the ensembles has almost identical overlap distribution across the operators, with the hybrid operator having the predominant overlap. The ground states also have a comparable distribution across the three ensembles, though we note that the work here includes an additional operator whose orbital structure is of the form ${\mathrm D}^{[2]}_{\mathrm{L}=0,\mathrm{S}}$ that can be interpreted as an operator of additional width with respect to the $S$-type orbital ${\mathrm D}^{[0]}_{\mathrm{L}=0,\mathrm{S}}$, and therefore of the same orbital structure.

For the case of the $N$, the identification and ordering of the hybrid baryons is comparable for the heavier $a094m358$ ensemble and the anisotropic ensemble, but for out lighter $a094m278$ ensemble, that identification is less obvious in spite of having significant overlap from the hybrid operators. As in the case of the $\Delta$, there is consistency in the overlaps for the ground state and the first two excited states across all three ensembles.

\section{Conclusions}\label{sec:conclusion}
In this work, we have computed the positive-parity $\Delta$ and $N$ spectra using an isotropic clover action. Our results support the observations in earlier works at heavier pion masses, and using the anisotropic clover action at a coarser spatial lattice spacing, but finer temporal lattice spacing. In particular, we find that rotational symmetry is largely observed at the hadronic scale, enabling us to reliably identify the spins of the states through their predominant overlap of operators derived from continuum operators of definite spin. However, the most significant outcome of this work is that we find that the spectra exhibit a counting of states in line with that of the quark model, but with additional states that we can identify as "hybrid" in nature, with the gluonic degrees of freedom playing a structural role.  The means used to identify such hybrids through the predominant overlap of a class of "hybrid" operators, pioneered in ref.~\cite{Dudek:2012ag}, must inevitably raise the issue of the operator dependence of such an identification. Here we use a different action, with a different lattice spacing and essentially different interpolating operators implemented through the variation of the number of distillation eigenvectors. Thus the identification of hybrid-type states in the spectrum is indeed robust. 

This work has important limitations in its use of ``single-hadron" operators which do not fully capture the low-lying energy levels in the finite-volume spectrum. The next step in the investigation of the nature of ``hybrid" baryons would be to include the multi-hadron operators, and subsequently to compute the infinite-volume momentum-dependent phase shifts.  Such a study could also reveal the decay modes of such states, and indeed the first study of the decays modes of the exotic $1^{-+}$ hybrid has recently been performed~\cite{Woss:2020ayi}.  This is more computationally challenging for baryons than for mesons through the increased cost of Wick contractions, the scaling of the number of distillation eigenvectors with increasing volume, and the numerous final states to which they can decay. Nonetheless, the advent of the exascale era of computation makes such computations increasingly realizable.  Ultimately, a largely model-independent determination of the quark and gluon content of such resonances will be achieved by the probing of their structure through external currents, and the theoretical framework for such studies is an area of rapid development \cite{Briceno:2015csa, Briceno:2014uqa, Briceno:2015tza} and application~\cite{Stokes:2019zdd, Stokes:2020gsd}.

\section{Acknowledgments}\label{sec:acknowledgement}
We thank Jozef Dudek, Robert Edwards, Archana Radhakrishnan and Christopher Johnson for useful discussions, and for the use of the \verb+reconfit+ fitting package. This work is supported by the U.S. Department of Energy, Office of Science, Office of Nuclear Physics under contract DE-AC05-06OR23177. Computations for this work were carried out in part on facilities of the USQCD Collaboration, which are funded by the Office of Science of the U.S. Department of Energy. This work was performed in part using computing facilities at The College of William and Mary which were provided by contributions from the National Science Foundation (MRI grant PHY-1626177), and the Commonwealth of Virginia Equipment Trust Fund. This work used the Extreme Science and Engineering Discovery Environment (XSEDE), which is supported by National Science Foundation grant number ACI-1548562. Specifically, it used the Bridges system, which is supported by NSF award number ACI-1445606, at the Pittsburgh Supercomputing Center (PSC) \cite{6866038, Nystrom:2015:BUF:2792745.2792775}. In addition, this work used resources at NERSC, a DOE Office of Science User Facility supported by the Office of Science of the U.S. Department of Energy under Contract \#DE-AC02-05CH11231, as well as resources of the Oak Ridge Leadership Computing Facility at the Oak Ridge National Laboratory, which is supported by the Office of Science of the U.S. Department of Energy under Contract No. \mbox{\#DE-AC05-00OR22725}. The software codes {\tt Chroma} \cite{Edwards:2004sx}, {\tt QUDA} \cite{Clark:2009wm, Babich:2010mu} and {\tt QPhiX} \cite{QPhiX2} were used in our work. The authors acknowledge support from the U.S. Department of Energy, Office of Science, Office of Advanced Scientific Computing Research and Office of Nuclear Physics, Scientific Discovery through Advanced Computing (SciDAC) program, and of the U.S. Department of Energy Exascale Computing Project. TK was support in part by the Center for Nuclear Femtography grants C2-2020-FEMT-006, C2019-FEMT-002-05.

\bibliographystyle{unsrt}

\bibliography{paper}
\end{document}